Causes of an AD 774-775 $^{14}$C increase

**Atmospheric $^{14}$C production is a potential window into the energy of solar proton and other cosmic ray events. It was previously concluded that results from AD 774-775 are orders of magnitude greater than known solar events[1]. We find that the coronal mass ejection energy based on $^{14}$C production is much smaller than claimed, but still substantially larger than the maximum historical Carrington Event of 1859[2-4]. Such an event would cause great damage to modern technology[5-6], and in view of recent confirmation of superflares on solar-type stars[7-8], this issue merits attention.**

It was computed[1] that for a solar proton event (intercepted by the Earth) to account for the $^{14}$C data, ~8 x $10^{18}$ J of kinetic energy would be needed at the Earth, about 20 times that estimated[3,4] for the historical Carrington Event of 1859. This was followed by an estimate of 2 x $10^{28}$ J for the originating coronal mass ejection. As this is several orders of magnitude beyond the range[9] of solar events, it was concluded that solar causation not reasonable.

The fluence for this event at the Earth based on $^{14}$C production[1] scaling to the coronal mass ejection energy was incorrect. The applied geometrical factor was the ratio of the area of a sphere with radius the Earth's orbit to the cross-sectional area of the Earth. This would be appropriate if coronal mass ejections propagated isotropically. Instead, the opening angles are tyically[10] 24° to 72°, with smaller angles much more common. We use 24°, corresponding to 0.01 of the surface area of a sphere. The implied energy of the coronal mass ejection is now reduced to ~2 x$10^{26}$ J.

The correct scaling puts solar proton event energy at the lower end of observed solar-type star superflare energies[7-8], which ranges from $10^{26}$-$10^{33}$ J, so there are solar-type stars with energy available to push coronal mass ejections well beyond the AD 774-775 event. Given poor constraints on the rates of such large events at the Sun[2,9], it would be wise to consider the possibility. A Carrington-level event would be disastrous for electromagnetic technology[5-6], causing widespread damage to satellites and transformers linking the power grid. No assessment has been made of the technological effects of an event 20 times stronger.

Solar flare rates follow a power-law in energy empirically based only for energies smaller than Carrington. The probability of the implied event can be studied using the statistics of rare events[11]. Using the information that there was just one such event within ~1250 years, the best estimate of the rate is 8 x $10^{-4}$ yr$^{-1}$, with 2σ (95.4%) confidence intervals from vanishingly small to 3.2 x $10^{-3}$ yr$^{-1}$. Therefore the estimate of the probability of such an event within the next decade is 0.8%, which may be viewed as small unless the devastating technological consequences are considered.

Based on rate/energy scalings we have examined previously[2], a short gamma-ray burst could cause such effects from within ~1 kpc, but with an a priori probability of order $10^{-4}$ over 1250 years.

Atmospheric ionization depletes ozone, increasing the solar UVB that reaches the ground[2,3]. We compute the ozone depletion[12] with corrected fluence and results shown in Figure 1. This is not a mass extinction level event; the results are consistent with moderate biological effects: reduction of primary photosynthesis in the oceans and increased risk of erythema and skin cancer, but no major mass-extinction level effects as implied earlier[1]. A newly published study[13] confirms our past computations of ozone depletion from a Carrington-level event, and suggests significant climate cooling might be a side effect, which would be enhanced for the event we describe here.

It is worth noting that the $^{14}C$ data could have been initiated by a series of solar proton events, each contributing a somewhat smaller amount. The $^{14}C$ production would add nearly linearly.

After these estimates were made, we noted a new study[14] of upper limits on energetic events at the Sun. An event of the energy resulting from our scaling lies just at their upper limits for an event that might appear every thousand years or so.

Therefore a solar proton event appears to be a possible cause, which demands further exploration of a potential massive threat to modern civilization.

Adrian L. Melott[1] & Brian C. Thomas[2]
*1 Department of Physics and Astronomy, University of Kansas, Lawrence, Kansas 66045 USA E-mail: melott@ku.edu*
*2 Department of Physics and Astronomy, Washburn University, Topeka, Kansas 66621 USA E-mail: brian.thomas@washburn.edu*

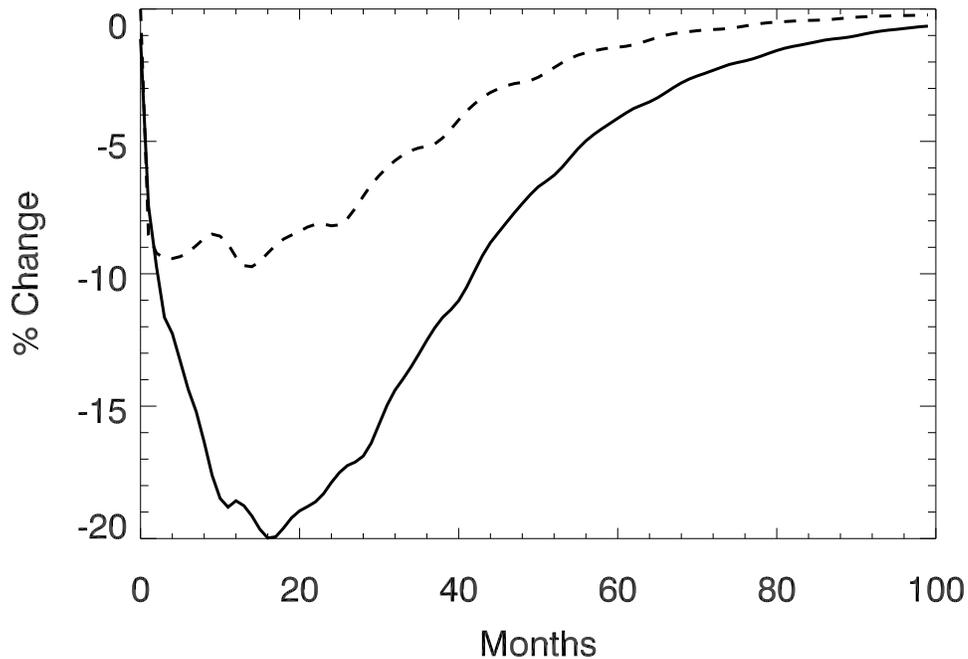

**Figure 1 caption:**
Percent change (comparing simulation runs with and without ionization input) in globally averaged $O_3$ column density. Dashed line = short GRB case; solid line = solar proton event case.

**Author Contributions**

ALM planned and wrote the communication with the assistance of BCT. BCT performed the atmospheric computations and made the plots.

**Author Information**

The authors declare no competing financial interests. Correspondence should be directed to A.L.M. (melott@ku.edu).